\documentclass[12pt,showpacs,prd]{revtex4}
\usepackage{epsfig,amssymb,amsfonts}
\begin{document}
\title{Field theory description of vacuum replicas}
\author{A. V. Nefediev}
\affiliation{Grupo Te\'orico de Altas Energias (GTAE), Centro de F\'\i sica das Interac\c c\~oes 
Fundamentais (CFIF),
Departamento de F\'\i sica, Instituto Superior T\'ecnico, Av. Rovisco Pais, P-1049-001 Lisboa, Portugal}
\affiliation{Institute of Theoretical and Experimental Physics, 117218,\\ B.Cheremushkinskaya 25, 
Moscow, Russia}
\author{J. E. F. T. Ribeiro}
\affiliation{Grupo Te\'orico de Altas Energias (GTAE), Centro de F\'\i sica das 
Interac\c c\~oes Fundamentais (CFIF),
Departamento de F\'\i sica, Instituto Superior T\'ecnico, Av. Rovisco Pais, P-1049-001 Lisboa, Portugal}
\newcommand{\be}{\begin{equation}}
\newcommand{\bea}{\begin{eqnarray}}
\newcommand{\ee}{\end{equation}}
\newcommand{\eea}{\end{eqnarray}}
\newcommand{\ds}{\displaystyle}
\newcommand{\low}[1]{\raisebox{-1mm}{$#1$}}
\newcommand{\loww}[1]{\raisebox{-1.5mm}{$#1$}}
\newcommand{\lmn}{\mathop{\sim}\limits_{n\gg 1}}
\newcommand{\vpint}{\int\makebox[0mm][r]{\bf --\hspace*{0.13cm}}}
\newcommand{\too}{\mathop{\to}\limits_{N_C\to\infty}}
\newcommand{\vp}{\varphi}

\begin{abstract}
In this paper we develop a systematic quantum field theory based approach
to the vacuum replica recently found to exist in effective low energy models in hadronic 
physics. A local operator creating the replica state is constructed explicitly. We
show that a new effective quark-quark force arises in result of replica existence. 
Phenomenological implications of such a force are also briefly discussed.
\end{abstract}
\maketitle

\section{Introduction}

In this paper we return to the possibility of vacuum replication put forward 
in a recent work \cite{BNR}. There it was suggested that at least one 
vacuum replica could exist in QCD and an explicit example of such a replica 
was produced for a general class of quark kernels compliant with chiral symmetry 
(see also \cite{oni} where a similar conclusion was made). 
Here we focus on the problem of quantum field theory formulation of hadronic processes 
involving such replicas. The paper is organized as follows: In the second section, 
as an introduction to the issue of vacuum replication, we give some details of the chiral 
model used to find such replicas, although, as we have already said, the existence of 
such states does not depend on a given particular model, but seems to be 
a general feature, found at least on two different approaches. In any case, the 
formalism we propose to take into account the effect of such replicas in hadronic processes 
is independent of any particular model --- any model dependence is stored in the chiral angle ---
and therefore we will not dwell too much on model details. In the third section, we briefly 
discuss how the vacuum replica appears in the theory, how it depends on the dimensionality 
of the space-time and on the strength and the form of the inter-quark interaction. 
We follow the original paper \cite{BNR}. The orthogonality and other properties of various vacuum 
states are the subject of the fourth section. Section five contains the new formalism 
necessary to describe the effect of the replica's new degree of freedom in
hadronic processes. Then, in the sixth section, we study propagation of quarks in presence of 
the replica, deriving and solving a Dyson-Schwinger equation for the single-quark Greens function. 
Overview of the results, as well as of the perspectives of the suggested method, 
are the subject of the concluding section of the paper. If not stated otherwise, 
we consider the chiral limit throughout the paper.

\section{The chiral Hamiltonian and the Bogoliubov-Valatin transformation}

\begin{table}[t]
\caption{Parameters of the model (\ref{H}) taken from the paper
\cite{BNR}.}
\begin{ruledtabular}
\begin{tabular}{cccc}
$\sigma_0$, $GeV^2$&$\alpha_s$&$U$, $MeV$&$\Lambda$, $MeV$\\
0.135&0.3&220&250\\
\end{tabular}
\end{ruledtabular}
\end{table}

In this section, we introduce the model and briefly outline the basic ideas needed 
for the remainder of the paper. We start from the chirally symmetric Hamiltonian,
\bea
\label{H}
&&H=\int d^3 x[H_0(\vec{x})+H_I^{(1)}(\vec{x})+H_I^{(2)}(\vec{x})],\nonumber\\
&&H_0(\vec{x})=\psi^\dagger(\vec{x},t)\left(-i\vec{\alpha}\cdot\vec{\bigtriangledown}\right)\psi(\vec{x},t),
\\
&&H_I^{(1)}(\vec{x})=\frac12\int d^3
y\;\psi^\dagger(\vec{x},t)\frac{\lambda^a}{2}\psi(\vec{x},t)\;
\biggl[V_0(|\vec{x}-\vec{y}|)+V_1(|\vec{x}-\vec{y}|)\biggr]\;\psi^\dagger(\vec{y},t)
\frac{\lambda^a}{2}\psi(\vec{y},t),\nonumber\\
&&H_I^{(2)}(\vec{x})=-\frac12\int d^3
y\;\psi^\dagger(\vec{x},t)\vec{\alpha}\frac{\lambda^a}{2}\psi(\vec{x},t)\;V_1(|\vec{x}-\vec{y}|)
\;\psi^\dagger(\vec{y},t)\vec{\alpha}\frac{\lambda^a}{2}\psi(\vec{y},t)\nonumber,
\eea
where $V_0(|\vec{x}|)=\sigma_0|\vec{x}|$, and
$V_1(|\vec{x}|)=-\frac{\alpha_s}{|\vec{x}|}\left(1-e^{-\Lambda
|\vec{x}|}\right)+U$. 
The Hamiltonian (\ref{H}) belongs to a certain
class of quark models having quark kernels ranging from delta-functional 
\cite{NJL} and
oscillator-type \cite{model,BR} to a more realistic linear form (see,
for example, \cite{linear}). Such models, hereafter referred as to 
NJL-type models, are very successful at describing 
low-energy hadronic properties, including scattering reactions like  $ \pi -\pi$ scattering 
\cite{pp}, where Ward identities {\em force} the cancellation between two types of scalar 
contributions: one arising from the exchange of the $q-\bar q$ bound-state sigma, which 
occurs naturally as a pole in the ladder summation for the scalar sector, and the other from 
the diffractive ({\it i.e.}, non-pole dominated) part of the $\pi -\pi$ T-matrix. The Weinberg $\pi -\pi$ 
scattering lengths then come naturally as a consequence of this cancellation. This is a general 
feature for hadronic scattering, solely governed by chiral symmetry invariance, 
and, therefore, not restricted to $\pi -\pi$ scattering or to {\em any particular kernel}. 
In fact the argument can be reversed to prove the need for the existence of sigma as a bound state of 
$q-\bar q$:  any microscopic theory with quarks, possessing 
chiral symmetry, {\bf must} contain an effective quark-quark sigma exchange, in order 
to {\bf cancel} the scalar contribution arising from the diffractive part (sometimes also 
termed as the \lq\lq contact term") of a given hadron-hadron scattering (like for instance $\pi -\pi$).  
It happens that the NJL-type model embodied in Eq.~(\ref{H}) possesses one vacuum replica. 
For an earlier derivation of the $\pi -\pi$ Weinberg scattering lengths, in the framework of contact 
quark kernels, both for the quartic fermion interaction and an $U(1)_A$-breaking six quark term, 
see ref.~\cite{pointlike}. Again, for that particular model, vacuum replicas were shown to occur.

{\em It is therefore remarkable that, across all these different quark kernels, the 
existence of vacuum replicas should constitute the rule rather than the exception}. 

On the other hand it is known \cite{cahill} that nonlocal-quark-kernel NJL models can be 
mapped --- via a bilocal bosonization --- onto a corresponding  effective chiral perturbation theory Lagrangian 
${\cal L}_{ChPT}$. Then from the above discussion on chiral symmetry invariance of NJL models and 
the role of the $\sigma$-meson, the probable nature of the physical $\sigma$ emerges: 
a bare $q-\bar q$ state highly distorted by $\pi -\pi$ 
coupled channels. So distorted that $\sigma$ \lq\lq doubling" may take place \cite{doubling}. 
The possible existence of a vacuum replica will further complicate this picture. It is the purpose 
of this paper to evaluate the contribution of such replicas to quark-quark kernels.

In what follows it is enough to assume the existence of one such replica, {\em without needing 
to resort to any particular model}, to be able to work out the consequences of replica existence 
for hadronic processes. However, in order to illustrate some of the concepts which will be needed, 
it is convenient to work within a particular model, which we choose to be the one defined in 
Eq. (\ref{H}). As we shall see, the results of this paper will be formally independent of this choice, 
with the model dependence encapsulated in one single function: $f(\omega )$.

The four parameters fixing the model (\ref{H}) are the string tension $\sigma_0$, the strong
coupling constant $\alpha_s$, the ultraviolet cut-off $\Lambda$ and the constant
potential $U$ which was used to fit for the right value of the chiral
condensate. It was argued in \cite{BNR} that with the choice of these parameters given
in Table I the model is governed by only one scale of order $300MeV$ and it naturally possesses 
both the Coulomb limit for heavy quarks and the chiral limit for massless quarks. The chiral
condensate and the constituent quark mass take their standard values,
\be
\langle\bar{\psi}\psi\rangle=-(250MeV)^3,\quad m_{\rm eff}=
\mathop{\lim}\limits_{p\to 0}\sqrt{p^2+m(p)^2}\approx 230MeV,
\ee
where $m(p)$ is the effective dynamical quark mass connected to the dressed-quark
dispersive law. The Bogoliubov-Valatin method is used to diagonalize the
Hamiltonian in terms of the dressed quarks, and the
mass-gap equation for the chiral angle $\varphi(p)$,
\be
\frac{\delta {\cal E}_{\rm vac}[\varphi]}{\delta\varphi(p)}=0,\quad
{\cal E}_{\rm vac}[\varphi]=\langle 0| TH[\varphi]|0\rangle,
\label{massgap}
\ee 
is studied in detail. There are many ways to define the chiral angle, and we
choose it to obey the following conditions \cite{BR}:
\be
-\frac{\pi}{2}\leqslant\vp(p)\leqslant\frac{\pi}{2},\quad \vp(0)=\frac{\pi}{2},\quad
\vp(p\to\infty)\to 0.
\ee

The dressed quark field is then
\be
\psi(\vec{x},t)=\sum_{\xi=\uparrow,\downarrow}\int\frac{d^3p}{(2\pi)^3}e^{i\vec{p}\vec{x}}
[b_{\xi}(\vec{p},t)u_\xi(\vec{p})+d_{\xi}^\dagger(-\vec{p},t)v_\xi(-\vec{p})],
\label{psi}
\ee
\be
\begin{array}{rcl}
u(\vec{p})&=&\frac{1}{\sqrt{2}}\left[\sqrt{1+\sin\vp}+
\sqrt{1-\sin\vp}\;(\vec{\alpha}\hat{\vec{p}})\right]u(0),\\
v(-\vec{p})&=&\frac{1}{\sqrt{2}}\left[\sqrt{1+\sin\vp}-
\sqrt{1-\sin\vp}\;(\vec{\alpha}\hat{\vec{p}})\right]v(0).
\end{array}
\label{uandv}
\ee 
\be
b_{\xi}(\vec{p},t)=e^{iE_pt}b_{\xi}(\vec{p},0),\quad
d_{\xi}(-\vec{p},t)=e^{iE_pt}d_{\xi}(-\vec{p},0),
\label{bandd}
\ee
with $E_p$ being the dressed-quark dispersive law. Notice a very important
property of the fermionic field (\ref{psi}). Formally it remains intact under
Bogoliubov-Valatin transformation with an arbitrary chiral angle $\vp(p)$ since
the rotation of the quark amplitudes $u$ and $v$ is compensated by
the contrarotation of the operators $b$ and $d$. Nevertheless, the
definition of the vacuum annihilated by the dressed single-particle operators, as
well as the definition of the one-particle state, change drastically, so that the
true vacuum, with the minimal vacuum energy, becomes full of strongly correlated
$^3P_0$ quark-antiquark pairs \cite{BR},
\be
|\tilde 0\rangle=S_0|0\rangle,\quad S_0=e^{Q_0^\dagger-Q_0},\quad
Q_0^\dagger=\frac12\int\frac{d^3p}{(2\pi)^3}\vp_0(p)C_p^\dagger,
\label{S0}
\ee
\be
C_p^\dagger=[b^\dagger_{\uparrow}(\vec{p}),b^\dagger_{\downarrow}(\vec{p})]
\mathfrak{M}_{^3P_0}
\left[\begin{array}{c}d^\dagger_{\uparrow}(\vec{p})\\
d^\dagger_{\downarrow}(\vec{p})\end{array}\right],
\label{Cddef}
\ee
with the $^3P_0$ matrix being
\be
\left[\mathfrak{M}_{^3P_0}\right]_{\xi_1\xi_2}=(-\sqrt{6})\sum_{\xi,m}
\left(\begin{array}{ccc}1&1&0\\m&\sigma&0\end{array}\right)
\hat{\vec{p}}_{1m}\left(\begin{array}{ccc}1/2&1/2&1\\
\xi_1&\xi_2&\sigma\end{array}\right)
=[(\vec{\sigma}\hat{\vec{p}})i\sigma_2]_{\xi_1\xi_2},
\label{mM}
\ee
where $\sigma$'s are the $2\times 2$ Pauli matrices.

The operator $C_p$ creates a $^3P_0$ quark-antiquark pair with zero total
momentum and the relative three-dimensional momentum $2\vec{p}$. The chiral
angle $\vp_0(p)$ is the solution to the mass-gap equation (\ref{massgap}) and it \lq\lq
measures" the weight of the pairs with the given  relative momentum, so
that the operator $S_0$ creates a cloud of correlated pairs, and the BCS
vacuum $|\tilde 0\rangle$ has the form of a coherent-like state. The dressed
fermionic propagator in this vacuum is given by the expression
\be
S(p_0,\vec{p})=\frac{\Lambda_+(\vec{p})\gamma_0}{p_0-E_p+i0}+
\frac{\Lambda_-(\vec{p})\gamma_0}{p_0+E_p-i0},
\label{S}
\ee
where the projectors $\Lambda_\pm$ are defined as 
\be
\Lambda_\pm(\vec{p})=\frac12[1\pm\gamma_0\sin\vp_0(p)
\pm(\vec{\alpha}\hat{\vec{p}})\cos\vp_0(p)].
\ee

A crucial statement which makes the model under discussion realistic is 
that, as said above, it is governed by only one scale, 
of order $250\div 300MeV$, which in turn 
can be dynamically generated, in a self-consistent way, with no 
other scales appearing (see Table I and the discussion in the reference \cite{BNR}).

\section{The replica}

The question whether the BCS vacuum possesses replicas was addressed in ref. \cite{BNR}. 
It was shown that the answer to this question strongly depended on the
dimensionality of the space-time, on the form of the potential, and on the strength
of the latter. For example, the two-dimensional model for QCD \cite{tHooft} was
found to possess no replica states at all, regardless of the form of the
potential \cite{BNR}. On the contrary, the four-dimensional potential does support the existence of 
replicas, provided it is strong enough. It was demonstrated long ago
that the oscillator-type potential leads to an infinite tower of solutions to
the mass-gap equation \cite{Orsay2}. On the other hand, one can numerically check that the pure linear 
potential fails by a very short margin, to  be  sufficiently \lq\lq binding" to hold replicas --- in fact 
it seems to be sitting on the brink of possessing such replicas. Then, it was suggested that the addition 
of a Coulomb potential --- which is both required by phenomenology and by the scale self-consistency 
needed for the class of models we considered --- might do the trick and tip the balance over to the side of 
potentials sufficiently attractive to possess a replica. Indeed this was the case and 
the existence of one, and only one, replica was found for the realistic set of parameters given in Table I. 
The existence of one replica for this class of realistic and scale self-consistent kernels together 
with the accumulated evidence that vacuum replication is likely to occur in four dimensions 
\cite{BNR,oni,Orsay2}, 
gives us reasons to expect that such a vacuum replication may occur in any
realistic model for QCD and, ultimately, be a property of QCD itself. The profiles of the chiral angles 
for the trivial, ground-state, and replica solutions are given in Fig.~1. Notice that both nontrivial 
solutions {\em converge
to each other in the ultraviolet regime} although for large quark momenta --- where the 
physical Fock space must contain gluons and the physics is controlled by perturbative QCD --- this class 
of models ceases to be reliable. Nevertheless, they still contain the key property that for large momenta the 
quarks should become asymptotically free, and chiral symmetry should be restored, although questions about 
the nature of this phase transition cannot be answered within the framework of these 
effective models. On the other hand, in the low energy regime, these models can be shown to possess all 
kind of good properties and to be able to give a global understanding of low-energy 
hadronic phenomenology \cite{pp}. 

\begin{figure}[t]
\centerline{\epsfig{file=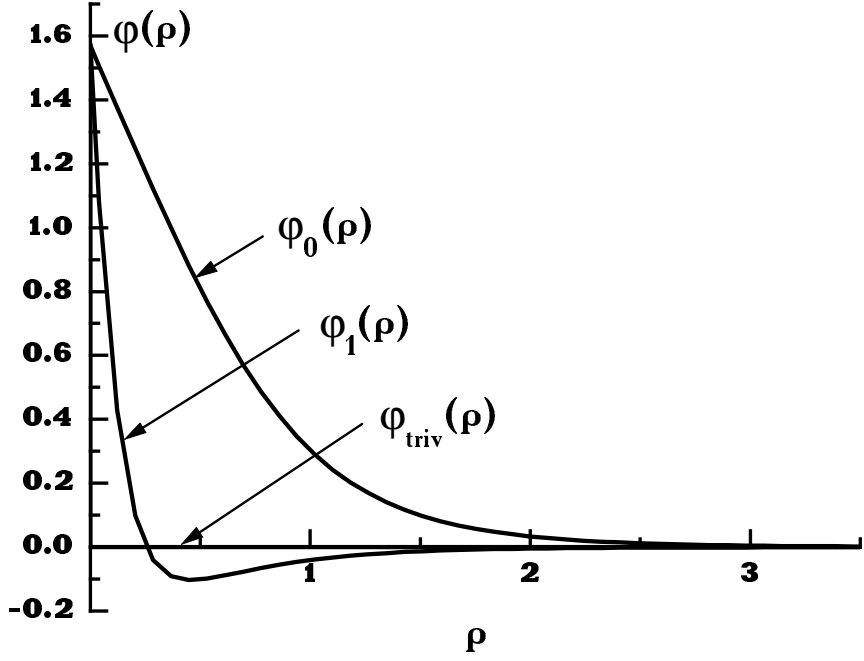,width=15cm}}
\caption{The three solutions to the mass-gap equation (\ref{massgap}): the trivial 
solution $\vp_{\rm triv}(p)=0$, the ground-state solution $\vp_0(p)$, and the replica 
$\vp_1(p)$. The momentum $p$ is given in the units of $\sqrt{\sigma_0}$ (see Table I).}
\end{figure}

{\em Therefore by vacuum replication we understand the phenomenon in which the 
same ultraviolet behaviour, for instance for the quark propagator, bifurcates to 
different solutions when we go to the low energy domain}. 

Due to the nonperturbative nature 
(and nonlinearity) of low energy QCD it is hard to exclude {\it a priori} this possibility which, in 
any case, is the rule rather than the exception for NJL-type models. 

\section{Properties of the vacuum}

\subsection{Orthogonality condition}

To start  let us
briefly overview the orthogonality properties of the BCS vacuum defined by 
Eq.~(\ref{S0}).

Using the commutation relations for the quark operators $b$ and $d$, 
one can easily find the following representation for the
ground-state vacuum \cite{BR}:
\be
|\tilde 0\rangle=\mathop{\prod}\limits_{p}\left(\cos^2\frac{\vp_0}{2}+
\sin\frac{\vp_0}{2}\cos\frac{\vp_0}{2}C^\dagger_p
+\frac12\sin^2\frac{\vp_0}{2}C^{\dagger 2}_p\right)|0\rangle\equiv S_0|0\rangle.
\label{nv}
\ee

If the trivial vacuum is normalized, $\langle 0|0\rangle=1$, then 
\be\label{norm00} 
\langle \tilde 0|\tilde 0\rangle=1,\quad \langle 0|\tilde 0\rangle=\prod_p
\cos^2\frac{\vp_0}{2}=\exp\left[V\int\frac{d^3p}{(2\pi)^3}\ln\left(\cos^2\frac{\vp_0}{2}\right)\right]
\mathop{\longrightarrow}\limits_{V\to \infty}0,
\ee
where the substitution $\sum_p\to V\int\frac{d^3p}{(2\pi)^3}$ was used; $V$
being the three-dimensional volume. Therefore, the chirally nonsymmetric vacuum
state is orthogonal to the trivial vacuum in the limit of infinite volume of the space. 
As usual, the Fock space is built over the $ |\tilde 0\rangle$ vacuum by repeated 
application of quark/antiquark creation operators belonging to the set 
$\{ {\tilde b}^\dagger_{\uparrow}(\vec{p}), {\tilde b}^\dagger_{\downarrow}(\vec{p}), 
{\tilde d}^\dagger_{\uparrow}(\vec{p}),{\tilde d}^\dagger_{\downarrow}(\vec{p})\}$, defined as,
\be\label{cantranf}
\tilde b=S_0 b S_0^\dagger,\quad \tilde d=S_0 d S_0^\dagger.
\ee 
Let us denote this Fock space  by $ {\cal F}_{\tilde 0}$. 
Once we have $ {\cal F}_{\tilde 0}$, we can address more 
complicated questions, like, for instance, Bethe-Salpeter amplitudes 
for hadrons and scattering (see ref. \cite{pp} for an example and references therein). 

For the replica, a relation similar to (\ref{nv}) can be written, with the
solution of the mass-gap equation $\vp_0$ changed for $\vp_1$ (see Fig.~1),
\be
|\tilde 1\rangle=\mathop{\prod}\limits_{p}\left(\cos^2\frac{\vp_1}{2}+
\sin\frac{\vp_1}{2}\cos\frac{\vp_1}{2}C^\dagger_p
+\frac12\sin^2\frac{\vp_1}{2}C^{\dagger 2}_p\right)|0\rangle\equiv S_1|0\rangle.
\label{nv2}
\ee

Then the orthogonality of the two nontrivial vacuum states can be easily proved after
simple algebraic transformations,
\be
\langle \tilde 0|\tilde 1\rangle=\prod_p
\cos^2\frac{\vp_0-\vp_1}{2}=
\exp\left[V\int\frac{d^3p}{(2\pi)^3}\ln\left(\cos^2\frac{\Delta\vp}{2}\right)\right]
\mathop{\longrightarrow}\limits_{V\to \infty}0,
\label{norm}
\ee
where the function $\Delta\vp(p)$ is given in Fig.~2.

\begin{figure}[t]
\centerline{\epsfig{file=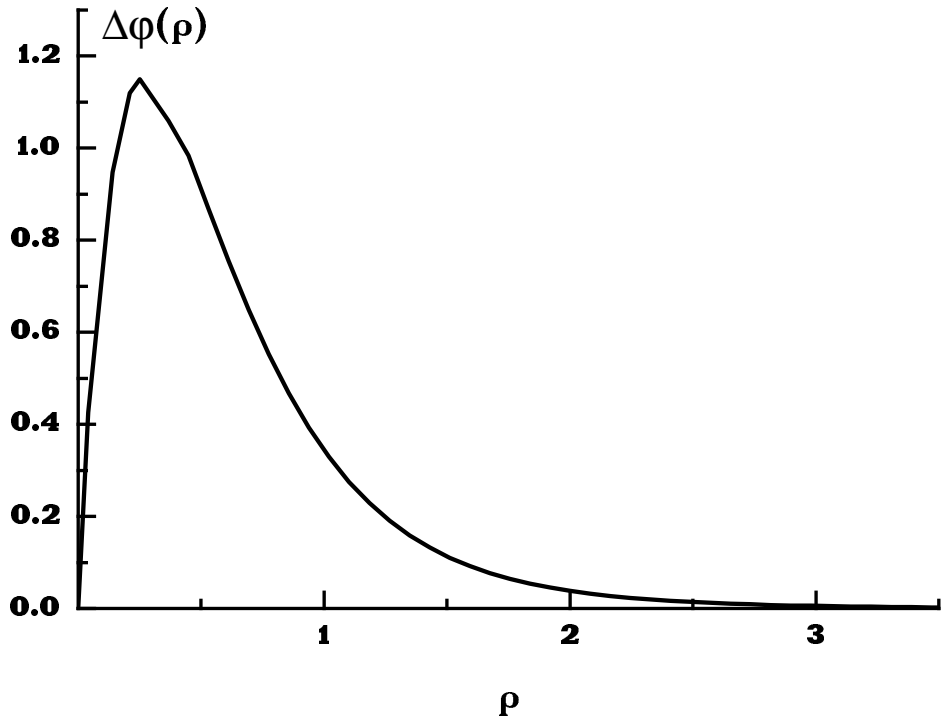,width=15cm}}
\caption{The difference of the ground-state and the replica solutions 
of the mass-gap equation. The momentum $p$ is given in the units of $\sqrt{\sigma_0}$
(see Table I).}
\end{figure}

Again, in the same fashion as we did for ${\cal F}_{\tilde 0}$, we can construct a 
Fock space over $|\tilde 1\rangle $, ${\cal F}_{\tilde 1}$. The difference is that 
the creation operators should be given by
\be\label{cantranf2}
\tilde b=S_1 b S_1^\dagger,\quad \tilde d=S_1 d S_1^\dagger,
\ee
instead of Eq.~(\ref{cantranf}). 

Let us study the unitary operator connecting the ground-state and the replica
vacua:
\be
|\tilde 1\rangle=S_1|0\rangle=S_1S_0^\dagger S_0|0\rangle=S_1S_0^\dagger 
|\tilde 0\rangle.
\label{op1}
\ee

Notice that, in order to have everything self-consistent, only operators inherent to the
ground-state vacuum must be used in Eq.~(\ref{op1}). For this purpose, we use the standard
transformation rules for the operators $b$ and $d$ in different representations to find
\be\label{umoper}
S_0=\tilde S_0^\dagger\tilde S_0\tilde S_0=\tilde S_0,\quad 
S_1=\tilde S_0^\dagger\tilde S_1\tilde S_0,
\ee
where the tilde denotes operators in the ground-state vacuum representation, as
opposed to the operators defined in the trivial vacuum, which do not have tildes.

Thus, Eq.~(\ref{op1}) can be rewritten as 
\be
|\tilde 1\rangle=\tilde S_0^\dagger\tilde S_1|\tilde 0\rangle=e^{\Delta\tilde
Q^\dagger-\Delta\tilde Q}|\tilde 0\rangle\equiv \tilde S|\tilde 0\rangle,
\label{op2}
\ee
where, similarly to Eq.~(\ref{S0}), we have,
\be
\Delta\tilde Q^\dagger=\frac12\int\frac{d^3p}{(2\pi)^3}\Delta\vp\tilde C_p^\dagger,
\label{dq}
\ee 
and all operators on the r.h.s. are appropriate for the ground-state vacuum. 

There is a crucial difference between the two Fock spaces. The ket $|\tilde 1\rangle $ does not 
correspond to the lowest energy state, which is $|\tilde 0\rangle $. In fact $|\tilde 1\rangle $ 
has an energy 
excess of $\Delta{\cal E}_{\rm vac}$ per volume unit \cite{BNR}. On the other hand, 
in the representation ${\cal F}_{\tilde 1}$, $|\tilde 0\rangle $ must correspond --- see 
Eqs. (\ref{op1}), (\ref{umoper}) --- to a 
coherent-like state, which entails an infinite superposition of quark-antiquark
pairs produced by quark/antiquark creation operators {\em consistent within 
that ${\cal F}_{\tilde 1}$ Fock space}. 
In general, for a given Hamiltonian, such a state would have a higher energy than the 
state without quark-antiquark pairs, which in ${\cal F}_{\tilde 1}$ is $|\tilde 1\rangle $, 
unless we have for 
that particular Hamiltonian, in this particular Fock space ${\cal F}_{\tilde 1}$, a tachyon. 
This is precisely the case. It is a simple consequence of the 
general Gell-Mann-Oakes-Renner relation \cite{GOR} and the fact that the chiral condensate 
in ${\cal F}_{\tilde 1}$ was found in \cite{BNR}
to have an opposite sign as compared to the chiral condensate in the BCS ($ {\cal F}_{\tilde 0}$) 
vacuum. Then, in the PCAC regime, whereas the lowest
state in the spectrum, the chiral pion, becomes  slightly massive, 
the mass of its replica  becomes imaginary, signaling the 
metastable nature of $|\tilde 1\rangle $. In the remainder of this paper we shall use the 
Fock space ${\cal F}_{\tilde 0}$ --- where the energy ordering of the states is natural, 
with $|\tilde 0\rangle$ being the lowest energy state, that is, the vacuum --- 
and will show $|\tilde 1\rangle$ to be a coherent state. 
From now on, the trivial vacuum will no longer be used, so that we
simplify notations omitting tildes and using $|0\rangle$ and $|1\rangle$ for the BCS
vacuum and the replica, respectively. If not stated explicitly, all field operators
are defined for ${\cal F}_{\tilde 0}$. 

With the help of Eq.~(\ref{op2}), it is straightforward to reproduce the normalization
condition (\ref{norm}) since, in analogy with (\ref{nv}), one has
\bea
\label{norm2}
&&\langle 0|1\rangle=\langle 0|
\mathop{\prod}\limits_{p}\left(\cos^2\frac{\Delta\vp}{2}+
\sin\frac{\Delta\vp}{2}\cos\frac{\Delta\vp}{2}C^\dagger_p
+\frac12\sin^2\frac{\Delta\vp}{2}C^{\dagger 2}_p\right)|0\rangle
\\
&&=\exp\left[V\int\frac{d^3p}{(2\pi)^3}\ln\left(\cos^2\frac{\Delta\vp}{2}\right)
\right]\approx
1-V\int\frac{d^3p}{(2\pi)^3}\left(\frac{\Delta\vp}{2}\right)^2+\ldots\nonumber,
\eea
where, for future references, we expanded the general expression for small
$\Delta\vp$. Notice that we can easily include an arbitrary number of colours and flavours by the 
simple substitution:
$$
C_p\to C_{pcf},\quad \sum_p\to\sum_{pcf},\quad \prod_p\to\prod_{pcf},
$$
with $c=1\ldots N_C$ and $f=1\ldots N_f$ being the colour and the
flavour indices, respectively. Then, the orthogonality condition
(\ref{norm2}) can be written in the full form as 
\be
\langle 0|1\rangle=\exp\left[N_CN_fV\int\frac{d^3p}{(2\pi)^3}\ln\left(\cos^2\frac{\Delta\vp}{2}\right)
\right]=\exp\left[{\rm Sp}\ln(1-\Gamma)\right],
\label{norm25}
\ee
where, in the last formula, we used the notations Sp$=\sum_{pcf}$,
$\Gamma=\sin^2\frac{\Delta\vp}{2}$ to arrive at the expression reminiscent of 
fermionic pair creation in the presence of external fields --- see Eq.~(4.84) of the ref. \cite{tb}. 
In other words, the scalar product $\langle 0|1\rangle$ can be written in the form of the matrix element 
$\langle 0|S|0\rangle$ with the operator $S[\psi,\bar\psi]=\Delta
Q^\dagger-\Delta Q$. In the next section, we discuss in detail the explicit form
of such an operator and an effective external field. 

\subsection{Coherent-like states and saturation of unity}

As mentioned above, the BCS vacuum as well as the replica have the form of
coherent-like states. The simplest example of a coherent state is provided by a
one-dimensional nonrelativistic harmonic oscillator with the mass $m$ and frequency 
$\omega$. It reads:
\be
|z\rangle=\frac{1}{\sqrt{2}}e^{\frac12|z|^2}e^{za^\dagger}|0\rangle,
\label{coh}
\ee
with $|0\rangle$ being the oscillator ground state --- the vacuum, and
$$
a^\dagger=\frac{1}{\sqrt{2}}\left(x\sqrt{m\omega}-i\frac{p}{\sqrt{m\omega}}\right),
$$
where $x$ and $p$ are the coordinate and the momentum operators, respectively.
The complex number $z$ parametrizes the coherent state which is
normalized, $\langle z|z \rangle=1$. 
Although coherent states are not orthogonal to each other,
\be
\langle z|z' \rangle=\exp\left(-\frac12|z|^2+z^*z'-\frac12|z'|^2\right),
\ee
they form an overcomplete set and it is possible to resolve the unity operator 
in terms of coherent states. In the simple case of the harmonic oscillator, we have
\be
\hat{I}=\int|z\rangle\frac{d^2z}{\pi}\langle z|=\int|z\rangle\frac{dzdz^*}{2\pi i}\langle z|,
\label{J1}
\ee
with a simple Jacobean $J(z,z^*)=1/2\pi i$. This  property of coherent states allows us
to use them as an alternative basis of states, which is known to be more appropriate 
than the one-particle basis in many applications. The problem of
propagation of coherent states is discussed in detail in \cite{cohs}. The state (\ref{op2}) constitutes 
yet another example of a coherent state. To construct the set of such coherent states, we use 
the property of the operators $C_p$:
$C_p^{\dagger n}|0\rangle=0$ for $n>2$, which in turn leads to the compact nonexponential form 
of the operator $S$ (see, for example, Eq.~(\ref{norm2})) \cite{BR}. Then, for each value of the
momentum $p$, the corresponding Fock space is spanned by three 
orthogonal states, $\{|0\rangle, C_p^\dagger[\theta_p]|0\rangle, C_p^{\dagger 2}[\theta_p]|0\rangle\}$, 
prompting us to generalize the operator $C_p^\dagger$ of equation 
(\ref{Cddef}) to a new operator $C_p^\dagger[\theta_p]$, defined as
\bea
&&C_p^\dagger[\theta_p]=[b^\dagger_{\uparrow}(\vec{p}),b^\dagger_{\downarrow}(\vec{p})]
\mathfrak{M}[\theta_p]
\left[\begin{array}{c}d^\dagger_{\uparrow}(\vec{p})\\
d^\dagger_{\downarrow}(\vec{p})\end{array}\right],\nonumber\\
&&\mathfrak{M}[\theta_p]=e^{i(\vec{\sigma}\hat{\vec{p}})\theta_p}\mathfrak{M}_{^3P_0},
\label{mnew}
\eea
where, according to Eq.~(\ref{mM}), the scalar coupling $^3P_0$ matrix is just
$\mathfrak{M}_{^3P_0}=(\vec{\sigma}\hat{\vec{p}})i\sigma_2$. One can easily check that an analogous matrix
for pseudoscalar $^1S_0$ pairing can be chosen to be $\mathfrak{M}_{^1S_0}=i\sigma_2$. It is the 
spin Salpeter amplitude, in the spin representation, for the pion \cite{BR}. 
Therefore the relation (\ref{mnew}) can be presented as
\be
\mathfrak{M}[\theta_p]=\mathfrak{M}_{^3P_0}\cos\theta_p+i\mathfrak{M}_{^1S_0}\sin\theta_p,
\ee
so that the angle $\theta_p$ defines rotations in the two-dimensional plane between the
scalar and pseudoscalar quark-antiquark couplings. One can easily check that the set of states
\be
|\alpha,\theta\rangle=
\mathop{\prod}\limits_{p}\left(\cos^2\frac{\alpha_p}{2}+
\sin\frac{\alpha_p}{2}\cos\frac{\alpha_p}{2}C^\dagger_p[\theta_p]
+\frac{1}{2}\sin^2\frac{\alpha_p}{2}C^{\dagger 2}_p[\theta_p]\right)|0\rangle
=\mathop{\prod}\limits_{p}|\alpha,\theta\rangle_p
\label{coh2}
\ee
with $-\pi<\alpha_p\leqslant \pi$ and $0\leqslant\theta_p<2\pi$ is complete, that is, it is 
possible to find at least one  Jacobean,
\be
J(\alpha_p,\theta_p)=\frac{1}{4\pi^2}\left(3-2\cos 2\alpha_p\right),
\label{J2}
\ee
as the simplest possibility, to saturate the unity operator, in the given space,
\be
\hat{I}=\int d[\alpha]d[\theta]|\alpha,\theta\rangle J\langle \alpha,\theta|
\equiv\mathop{\prod}\limits_{p}\int_{-\pi}^{\pi} d\alpha_p\int_0^{2\pi}d\theta_p
{|\alpha,\theta\rangle}_p J(\alpha_p,\theta_p)_p\langle\alpha,\theta|.
\label{unit}
\ee

Similarly to the case of coherent states (\ref{coh}), the choice (\ref{J2}) 
is not unique. For example,
$$
J'(\alpha_p,\theta_p)=J(\alpha_p,\theta_p)+\sum_{n=3}^{\infty}A_n\cos(n\alpha_p)
+\sum_{n=1}^{\infty}B_n\sin(n\alpha_p)
$$
with arbitrary coefficients $A_n$ and $B_n$ also supports the condition (\ref{unit}).
The replica $|1\rangle $ corresponds to a specific choice of the parameters, 
$\alpha_p=\Delta\vp(p)$ and $\theta_p=0$. The Fock space built in the usual way over the 
set $|\alpha,\theta\rangle$ constitutes an example of an overcomplete space.

\section{Quantum field theory based approach to the replica}

In this section, we introduce a systematic approach to the replica state, 
which can be used to investigate the contribution of the excited vacuum bubbles
to hadronic processes.

\subsection{Replica creator as a local operator}

Using the explicit form of the quark field (\ref{psi}), one can find that $\Delta
Q^\dagger$ defined in (\ref{dq}) is expressed as
\be
\begin{array}{rcl}
\Delta Q^\dagger&=&+\int d^4x
f(x_0)\bar\psi(\vec{x},x_0)\left(-i\vec{\alpha}\cdot\vec{\bigtriangledown}\right)
\Lambda_-\psi(\vec{x},x_0),\\
\Delta Q&=&-\int d^4x
f(x_0)\bar\psi(\vec{x},x_0)\left(-i\vec{\alpha}\cdot\vec{\bigtriangledown}\right)
\Lambda_+\psi(\vec{x},x_0),
\end{array}
\label{f0}
\ee  
if the function $f(x_0)$ is even with the Fourier transform being even,
$f(\omega)=f(-\omega)$, and
\be
f(2E_p)=\frac{\Delta\vp(p)}{2p}.
\label{f}
\ee
To exemplify the form of the function $f(\omega)$, let us use a simple 
anzatz of the dispersive law, 
\be
E_p=\frac{\vec{p}^2}{|\vec{p}|+\Delta}-\Delta,
\label{dl1}
\ee
which begins at $E_p=-\Delta$ $(\Delta>0)$ when $\vec{p}=0$ and tends to the free limit, 
$E_p=|\vec{p}|$, when $|\vec{p}|\to\infty$. The parameter $\Delta$ is
generated nonperturbatively and it has the order of the interaction scale
(see the discussion in refs. \cite{model,BR} and, for the two-dimensional QCD, 
in ref. \cite{2d}). The appropriate solution of the equation $E_p=\omega/2$ 
is found analytically to be,
\be\label{analitica}
p_*(\omega)=\frac14\left[\omega+2\Delta+\sqrt{(\omega+2\Delta)(\omega+10\Delta)}\right],
\ee  
if $\omega\geqslant -2\Delta$, and no solution exists otherwise. Then the function 
$f(\omega)$ takes the form
\be
f(\omega)=\frac{\Delta\vp(p_*(\omega))}{2p_*(\omega)}\theta(2\Delta+\omega)+
\frac{\Delta\vp(p_*(-\omega))}{2p_*(-\omega)}\theta(2\Delta-\omega),
\label{ffit1}
\ee
with $\theta$ being the Heaviside step-like function.
Now we fit the curve for $\Delta\vp(p)$ in Fig.~2,
\be
\Delta\vp(p)=2Lp\exp\left(-\frac{p}{\lambda}\right),\quad
L\approx\frac{5.6}{\sqrt{\sigma_0}},\quad\lambda\approx 0.29\sqrt{\sigma_0},
\label{fit1}
\ee
and substitute it into Eq.~(\ref{ffit1}) to obtain an expression for $f(\omega)$ which 
turns out to be a function localized in the regions
$|\omega|\approx 2\Delta\sim\sqrt{\sigma_0}$ with the smearing given by the parameter $\lambda$ (for the  value of $\sigma_0=0.135 GeV^2$ given in Table I, 
$\lambda\approx 100MeV$). This defines the most probable size of a $|1\rangle$-bubble to be 
$\nu\sim\sqrt{\sigma_0}/\Delta{\cal E}_{\rm vac}$. Notice that since the volume density of the energy 
$\Delta{\cal E}_{\rm vac}$, stored in the bubble containing the replica, is also fixed by the same 
parameters of the interaction \cite{BNR}, then the distribution in energies can also be seen as a 
distribution in volumes,
with $f(\Delta{\cal E}_{\rm vac}\times \nu)$ playing the role 
of the weight of the contribution for each volume $\nu$, so that the volume of a
$|1\rangle$-bubble is fixed by the interaction scale and it is not a free parameter.
 
Using the relations (\ref{f0}), one can easily find the following 
representation for the operator $S=\exp[\Delta Q^\dagger-\Delta Q]$, which relates the
ground state and the replica \cite{mtk}:
\be
S_T=T\exp\left[\int d^4x
f(x_0)\bar\psi(\vec{x},x_0)\left(-i\vec{\alpha}\cdot\vec{\bigtriangledown}\right)
\psi(\vec{x},x_0)\right],
\label{Sop}
\ee
where the projector operators disappear due to the equality $\Lambda_++\Lambda_-=1$
and we used time ordering for noncommuting field operators. Eq.~(\ref{Sop}) is general, 
the model dependence encapsulated in $f(x_0)$. The difference of the chiral angles $\Delta\vp$ 
being a small localized function (see Fig.~2) allows the average 
$\int\frac{d^3p}{(2\pi)^3}\left(\frac{\Delta\vp}{2}\right)^2$ 
of Eq.~(\ref{norm2}) to be a perturbative expansion parameter so that, as a result, 
one can also use a perturbative expansion of $S_T$ in the number of insertions of the operator 
$(-i\vec{\alpha}\cdot\vec{\bigtriangledown})$ multiplied by the function
$f(\omega)$. In what follows, it is sufficient to truncate this perturbative series at the quadratic term. 
Contributions of higher order terms could be calculated in the same fashion.

\begin{figure}[t]
\begin{tabular}{ccc}
\epsfig{file=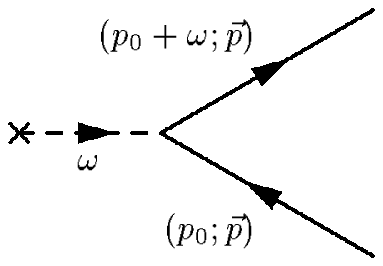,width=5cm}&\hspace*{2cm}&
\epsfig{file=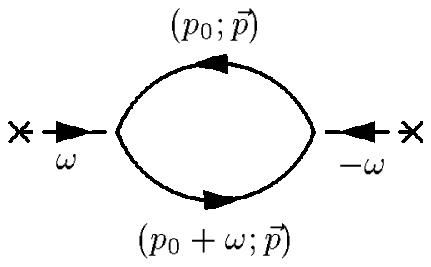,width=5cm}\\
(a)&&(b)
\end{tabular}
\caption{The f-vertex (diagram (a)) and an example of the diagram with two 
f-vertex insertions (diagram (b)), which gives the lowest contribution to the 
orthogonality condition (\ref{norm3}).}
\end{figure}

The local operator (\ref{Sop}) creating the replica vacuum state, together with 
the definition (\ref{f}), is the central object of this paper. It describes the interaction with an external 
field (see Fig.~3a) of a particular form whose physical meaning is local excitations of the BCS vacuum, that 
is, fluctuations  of the quark-antiquark correlations in the vacuum. In the remainder of the paper we call 
this vertex f-vertex.

The operators defined by Eqs.~(\ref{op2}) and (\ref{Sop}) are not
identical. The crucial difference between them comes from the time ordering which appears in the
definition (\ref{Sop}). Indeed, for Eq.~(\ref{op2}) and arbitrary time-dependent operators ${\cal O}_i(t)$
one has 
\be
\langle 0| T {\cal O}_1 (t_1)\ldots{\cal O}_n (t_n)S|0\rangle=
\langle 0|[T{\cal O}_1 (t_1)\ldots{\cal O}_n (t_n)]\sum_{n=0}^{\infty}\frac{1}{n!}
(\Delta Q^\dagger-\Delta Q)^n|0\rangle,
\label{av1}
\ee
where all operators creating the excited vacuum stand to the right from the
operators ${\cal O}$. On the contrary, with definition (\ref{Sop}), one has the possibility to create 
the replica state at intermediate moments of time:
$$
\langle 0| T {\cal O}_1 (t_1)\ldots{\cal O}_n (t_n)S_T|0\rangle\approx
\langle 0| {\cal O}_1 (t_1)\ldots{\cal O}_n (t_n)|0\rangle\hspace*{6cm}
$$
\be
+\int_{t_1}^{\infty}dtf(t)\langle 0| [\bar\psi(-i\vec{\alpha}\cdot\vec{\bigtriangledown})
\psi]_t{\cal O}_1 (t_1)\ldots{\cal O}_n (t_n)|0\rangle
\label{av2}
\ee
$$
\hspace*{5.5cm}+\int_{t_2}^{t_1}dtf(t)\langle 0| {\cal O}_1 (t_1)
[\bar\psi(-i\vec{\alpha}\cdot\vec{\bigtriangledown})
\psi]_t{\cal O}_2 (t_2)\ldots{\cal O}_n (t_n)|0\rangle+\ldots,
$$
where we choose $t_1>t_2>\ldots >t_n$.
This difference does not play a role if no external operators are involved, as in the case of the vacua 
orthogonality condition (see Eq.~(\ref{norm3}) below), but it becomes extremely important for the averages 
similar to
(\ref{av1}), (\ref{av2}). Thus, with the operator (\ref{Sop}), we extend the applications of the vacuum 
replica allowing its creation for finite periods of time during hadronic reactions.

\subsection{Vacua orthogonality revisited}

Let us demonstrate how the orthogonality condition (\ref{norm2}) appears in the new
field-theory approach. From (\ref{Sop}) we have,
\be
\langle 0|1\rangle=\langle 0|S|0\rangle =\langle 0|T\exp\left[\int d^4x
f(x_0)\bar\psi(\vec{x},t)\left(-i\vec{\alpha}\cdot\vec{\bigtriangledown}\right)
\psi(\vec{x},t)\right]|0\rangle
\label{norm3}
\ee
$$
\approx 1+\langle 0|\int d^4x
f(x_0)\left[\bar\psi\left(-i\vec{\alpha}\cdot\vec{\bigtriangledown}\right)
\psi\right]_x|0\rangle
$$
$$
+\frac12\int d^4 x d^4 y f(x_0)f(y_0)\langle 0|T
\left[\bar\psi\left(-i\vec{\alpha}\cdot\vec{\bigtriangledown}\right)\psi\right]_x
\left[\bar\psi\left(-i\vec{\alpha}\cdot\vec{\bigtriangledown}\right)\psi\right]_y|0\rangle
+\ldots,
$$
where, as explained before, we truncate the expansion at the level $f^2$. It is easy
to see that the term of order $f$ vanishes since 
$\bar v_\sigma(-\vec{p})(\vec{\alpha}\vec{p})v_{\sigma'}(-\vec{p})=0$, so that the
lowest contribution to (\ref{norm3}) is given by the quadratic term. Moreover, it is
clear that all terms with odd number of f-vertices vanish. Diagrammatically, the
series (\ref{norm3}) is given by the sum of many-leg loops with even number of
f-vertices (see the lowest term in Fig.~3b). We have,
\be
\langle 0|1\rangle=1+V\int\frac{d^4p}{(2\pi)^4}\frac{d\omega}{2\pi}f(\omega)f(-\omega){\rm Sp}
\left[(\vec{\alpha}\vec{p})S(p_0,\vec{p})(\vec{\alpha}\vec{p})
S(p_0+\omega,\vec{p})\right]+\ldots 
\label{norm4}
\ee
or, after substitution of the explicit form of the propagator $S(p_0,\vec{p})$ from 
Eq.~(\ref{S}) and straightforward algebraic calculations, 
\be
\langle
0|1\rangle=1-V\int\frac{d^3p}{(2\pi)^3}\left(\frac{\Delta\vp}{2}\right)^2+\ldots,
\label{norm5}
\ee
where the property (\ref{f}) of the function $f$ was also used. The expression (\ref{norm5})
coincides with the expansion in Eq.~(\ref{norm2}). To restore the full form of Eq. 
(\ref{norm2}), notice that the factor of the three-dimensional volume $V$ accompanies
each fermionic loop in the diagram. Thus, summing all one-loop many-leg diagrams one
restores the logarithm $\ln\left(\cos^2\frac{\Delta\vp}{2}\right)$. Now, to arrive at
the exponential form, we need only to take  into account many-loop disconnected diagrams
to reproduce the whole series of terms $\left[{\rm one\;loop}\right]^n/n!$. 
They sum to yield Eq.~(\ref{norm2}). The colour and the
flavour factors follow immediately from the corresponding traces over the fermionic
loops, so that one can easily see that they always accompany the volume $V$ and
enter in the combination $N_CN_fV\int\frac{d^3p}{(2\pi)^3}={\rm Sp}_{pcf}$.
This completes the proof and shows that $S_T$ is the quantum-field creator of the replica. 
It is clear now that normalization of the matrix element of
an arbitrary operator ${\cal O}$, in presence of the replica,
\be
\langle 0|T{\cal O}S_T|0\rangle\to\langle 0|T{\cal O}S_T|0\rangle_{\rm normalized}=
\frac{\langle 0|T{\cal O}S_T|0\rangle}{\langle 0|S_T|0\rangle}=\langle 0|{T\cal
O}S_T|0\rangle_{\rm connected},
\label{nop1}
\ee
means that only connected diagrams should be considered with Wick contractions
between fermionic operators coming from the operator ${\cal O}$ itself and those from
the operator $S_T$ given by Eq.~(\ref{Sop}).

\section{Feynman rules in the presence of the replica}

In this subsection, we address the question of the quark propagation in the vacuum in 
the presence of the replica.

From the operator (\ref{Sop}) one can
easily derive the Feynman rules for the f-vertex (Fig.~3a):
\begin{itemize}
\item each f-vertex is to be supplied by the product
$f(\omega)(\vec{\alpha}\vec{p})$, where $\omega$ is the energy pumped to the
system in this vertex, $\vec{p}$ being the three-dimensional momentum
floating into the f-vertex along the fermionic line;
\item for each f-vertex one has an integral over the energy $\omega$.
\end{itemize} 
However, as it stands, {\em global energy conservation is not satisfied} because 
the operator $S_T$ of Eq.~(\ref{Sop}) does not satisfy time translational invariance. 
To restore time translational invariance we are forced --- very much in the same way 
as it was done in the instanton case --- to consider here a one-dimensional gas
of replicas with each of them associated with a particular time origin.

\subsection{Time translational invariance and time chains of replicas: the f-chain}

The function $f(x_0)$ defined in Eqs.~(\ref{f}), (\ref{analitica}), and (\ref{ffit1}) 
describes a localized in time object with the centre at $x_0^{(0)}=0$. 
However, translational invariance in time would require results to be independent of 
the time origin. To achieve time translational invariance we shall proceed along similar 
lines as were used in the case of instantons. First, let us generalize the formulae
(\ref{f0})-(\ref{Sop}) to describe the f-vertex localized near a nonzero time origin 
$x_0^{(0)}$, $f(x_0-x_0^{(0)})$. Then, the modified version of the relation 
(\ref{ffit1}) takes the form
\be
f(\omega)=f^*(-\omega)=\left[\frac{\Delta\vp(p_*(\omega))}{2p_*(\omega)}\theta(2\Delta+\omega)+
\frac{\Delta\vp(p_*(-\omega))}{2p_*(-\omega)}\theta(2\Delta-\omega)\right]e^{i\omega x_0^{(0)}}
\label{f3}
\ee
with $p_*(\omega)$ being the same solution of the equation $E_p=\omega/2$ as in Eq.~(\ref{f}). 
The operator (\ref{Sop}) becomes
\be
S_T=T\exp\left[\int d^4x
f(x_0-x_0^{(0)})\bar\psi(\vec{x},x_0)\left(-i\vec{\alpha}\cdot\vec{\bigtriangledown}\right)
\psi(\vec{x},x_0)\right].
\label{Sop22}
\ee

Finally, it suffices to sum over an infinite set of time origins to restore time translations. As for this 
set, we would be forced to solve the dynamics of replica interactions, which is beyond the scope of this 
paper. As in the instanton case we assume a simpler picture of a dilute gas (in time) of replicas. So 
let us consider a diluted chain of $N$ time origins $\{f(x_0-x_0^{(n)})\}_N$, here called f-chain:
\be
F(x_0)=\sum_{n=1}^{N}f(x_0-x_0^{(n)})\equiv \sum_{n=1}^{N}f_n(x_0),
\label{Fdef}
\ee
\be
f_n(\omega)=\left[\frac{\Delta\vp(p_*(\omega))}{2p_*(\omega)}\theta(2\Delta+\omega)+
\frac{\Delta\vp(p_*(-\omega))}{2p_*(-\omega)}\theta(2\Delta-\omega)\right]e^{i\omega x_0^{(n)}}.
\label{fn}
\ee

Notice that for well separated
functions $f_n(x_0)$ in the sum (\ref{Fdef}) all interference terms are suppressed:
\be
f_n(\omega)f_m(\omega)\propto \delta_{nm},
\label{nrm}
\ee
for the separation $|x_0^{(n)}-x_0^{(m)}|\gg 1/\lambda$. This is the case for a diluted chain. As a result, 
we have the sum of $N$ equivalent contributions coming from $N$ independent isolated f-vertices. For $N$ 
going to infinity, with the fixed mean
separation between two neighbours $\tau$, one naturally arrives at the notion of the density $n=N/T= 1/\tau$. 
It is sufficient to consider an equally spaced one-dimensional f-chain with $x_0^{(n)}=n\tau$. Within this 
dilute approximation it is straightforward to recover the overall energy conservation law (time 
translational invariance). Indeed, if a 
diagram does not contain external legs, like the fish-like diagram in Fig.~3b, then the overall phase 
factor coming from the definition (\ref{f3}) disappears automatically due to the properties of the 
f-vertex (see Eq.~(\ref{fn})):
$$
f_n(\omega)f_n(-\omega)=f_n(\omega)f_n^*(\omega)=|f_n(\omega)|^2.
$$

On the contrary, in the case of $N_{\rm ext}$ external legs attached to the diagram 
we have a phase factor for the $m$th f-vertex in the chain, 
the exponent $\exp[ix_0^{(m)}\sum_{i=1}^{N_{\rm ext}}E_i^{\rm
ext}]$, where $E_i^{\rm ext}$ is the energy delivered by the $i$th external leg of 
the diagram of $N_{\rm ext}$. As a result, if the total energy is not conserved, 
$\sum_{i=1}^{N_{\rm ext}}E_i^{\rm ext}\neq 0$, then each f-vertex brings in a contribution to the total 
sum with a random phase factor, so that the total
amplitude vanishes. Formally, this can be demonstrated as follows:
\be
\sum_{m=1}^{N}\exp\left(ix_0^{(m)}\sum_{i=1}^{N_{\rm ext}}E_i^{\rm ext}\right)\approx
\sum_{m=1}^{N}\exp\left(im\tau\sum_{i=1}^{N_{\rm ext}}E_i^{\rm ext}\right)\mathop{\longrightarrow}
\limits_{N\to\infty}
\frac{1}{\tau}\delta\left(\sum_{i=1}^{N_{\rm ext}}E_i^{\rm ext}\right).
\label{econ}
\ee

The physical reason for such a relation is the fact that, for an infinitely long f-chain, one recovers the 
translational invariance in time and therefore the total energy conservation law is also restored. 

Then time translation invariance can be coded into two extra Feynman rules, in addition to the previous two 
rules:
\begin{itemize}
\item the full diagram is to be supplied by an overall energy conservation 
$\delta$-function, so that for the diagram with K f-vertices only K-1 independent 
integrals over pumped energies remain;
\item the full amplitude is to be multiplied by the time density of the replica 
excitations, $n=N/T$, which accounts for the fact that during the full time of the 
hadronic process $T$ the replica has been excited $N$ times.
\end{itemize}

\subsection{Fermionic propagator in the presence of replica} 

The quark propagator in the presence of the replica is given by the formula:
\be
S'(x)=\langle 0|T\psi(x)\bar\psi(0)S_T|0\rangle_c,
\label{Sx}
\ee
where $S_T$ is the operator creating the f-chain and, as discussed before, only
connected diagrams are to be considered. Expanding the propagator (\ref{Sx}) in powers of
f-vertex insertions with a truncation at the quadratic term, one finds:
$$
S'(x)\approx \langle 0|T\psi(x)\bar\psi(0)|0\rangle+\int d^4y
F(y_0)\langle 0|T\psi(x)\bar\psi(0)
\left[\bar\psi\left(-i\vec{\alpha}\cdot\vec{\bigtriangledown}\right)
\psi\right]_y|0\rangle
$$
\be
+\int d^4y d^4z F(y_0)F(z_0)
\langle 0|T\psi(x)\bar\psi(0)
\left[\bar\psi\left(-i\vec{\alpha}\cdot\vec{\bigtriangledown}\right)
\psi\right]_y\left[\bar\psi\left(-i\vec{\alpha}\cdot\vec{\bigtriangledown}\right)
\psi\right]_z|0\rangle+\ldots,
\label{Sx2}
\ee
with $F(t)$ being the f-chain profile function. In momentum space, using the Feynman
rules for the replica derived above, one arrives at
$$
S'(p_0,\vec{p})=S(p_0,\vec{p})+nf(0)S(p_0,\vec{p})(\vec{\alpha}\vec{p})S(p_0,\vec{p})
$$
\be
+n\int\frac{d\omega}{2\pi}f(\omega)f(-\omega)S(p_0,\vec{p})(\vec{\alpha}\vec{p})
S(p_0+\omega,\vec{p})(\vec{\alpha}\vec{p})S(p_0,\vec{p})+\ldots.
\label{Sp1}
\ee
We neglect all terms containing $f(0)$, since, in our model, $f(0)$ turns out to be very small 
$(f(0)\ll 1/\sqrt{\sigma_0})$ albeit not exactly zero. It corresponds to large quark momenta, 
$p\sim p_*(\omega=0)\sim 300 MeV$, and it is clear that for this range of quark momenta we should 
have only one phase --- governed by perturbative QCD with quarks {\em and gluons} --- and, 
therefore, $\Delta\vp$ should be identically zero rather than being very small. Having no gluons, 
being a mean-field theory, in the gluonic sector, the class of Nambu-Jona-Lasinio models we use in 
this paper cannot address the physics in this region,
being already remarkable that chiral symmetry requirements alone should force an almost identical 
asymptotic behaviour, in momenta, for the quark dispersive relations in different chiral phases.

Having this in mind, we finally arrive at simplified Feynman rules in the presence of replicas. They are:
\begin{itemize}
\item each f-vertex is supplied by the product $(\vec{\alpha}\vec{p})$, where $\vec{p}$ is the 
three-dimensional momentum floating into the f-vertex along the fermionic line;
\item the two f-vertices are connected by an effective f-propagator of the form
\be
G(p)=nf(p_0)f(-p_0)(2\pi)^3\delta^{(3)}(\vec{p})
\label{Gn}
\ee
with $n$ being the time density of f-vertices in the chain;
\item global conservation of the energy and the three-momentum is ensured in the usual way.
\end{itemize}
Notice that we chose to move the distribution function $f(\omega)$ from the vertex to 
the propagator in order to have its form close to those adopted in the solid-state physics.

Using the new Feynman rules, one can rewrite formula (\ref{Sp1}) for the propagator as a Dyson series,
\be
S'(p)=S(p)+S(p)\Sigma(p)S(p)+\ldots,
\label{Spr}
\ee
\be
\Sigma(p)=\int\frac{d^4q}{(2\pi)^4}G(p-q)(\vec{\alpha}\vec{q})iS(q)(\vec{\alpha}\vec{p}),
\label{sig}
\ee
where $S(p)$ is given by Eq.~(\ref{S}). Straightforward calculations yield:
\be
\Sigma(p)=\frac{n}{2}|\vec{p}|^2\left[\gamma_0\Lambda_+(\vec{p})|f(p_0+E_p)|^2-
\gamma_0\Lambda_-(\vec{p})|f(p_0-E_p)|^2\right],
\label{sig2}
\ee
and Eq.~(\ref{Spr}) can be approximately rewritten as
\be
S'(p)\approx\frac{\Lambda_+(\vec{p})\gamma_0}{p_0-E_p-\frac{n}{2}|\vec{p}f(p_0+E_p)|^2+i0}+
\frac{\Lambda_-(\vec{p})\gamma_0}{p_0+E_p+\frac{n}{2}|\vec{p}f(p_0+E_p)|^2-i0}.
\label{Spr1}
\ee

Using the fact that the first term in Eq.~(\ref{Spr1}) is peaked at $p_0\approx E_p$ and
the second term is peaked at $p_0\approx -E_p$, we can neglect the effect of the dressing on the 
projectors and approximately substitute $f(p_0\pm E_p)$ by $f(2E_p)$ in both 
denominators, which yields:
\be
S'(p)\approx
\frac{\Lambda_+(\vec{p})\gamma_0}{p_0-(E_p+\delta E_p)+i0}+
\frac{\Lambda_-(\vec{p})\gamma_0}{p_0+(E_p+\delta E_p)-i0}
\approx S(p)+\frac{\partial S(p)}{\partial E_p}\delta E_p+\ldots,
\label{Spr2}
\ee
where
\be
\delta E_p=\frac{n}{8}(\Delta\vp)^2.
\label{dE}
\ee

In other words, the effect of the replica on the fermionic propagator amounts to a shift 
of the poles of the latter from $\pm E_p$, as in Eq.~(\ref{S}), to $\pm[E_p+\frac{n}{8}(\Delta\vp)^2]$.

Relations (\ref{Spr}) and (\ref{sig}) together can be presented in the form of the Dyson-Schwinger equation 
for the new fermionic propagator in the presence of the replica:
\be
S'^{-1}(p)-S^{-1}(p)=i\int\frac{d^4q}{(2\pi)^4}
G(p-q)(\vec{\alpha}\vec{q})S'(q)(\vec{\alpha}\vec{p}),
\label{DSeq}
\ee
which we solve up to the terms of order $(\Delta\vp)^2$. We do not dress the vertex in Eq.~(\ref{DSeq}), 
since this leads to higher corrections in powers of $\Delta\vp$. Thus we conclude that, at the given level 
of accuracy, quarks propagate in the presence of the replica, as if a new type of inter-quark interaction 
defined by the quark current-current interaction kernel (\ref{Gn}) had been introduced. This {\it effective} 
interaction is present only in the BCS vacuum with spontaneously broken chiral symmetry and amounts to the 
change of the quark dispersive law, as given in Eq.~(\ref{dE}).

\section{Discussion and Conclusions}

References \cite{BNR} and \cite{oni}, when taken together, show that vacuum 
replication should constitute the rule rather than the exception for, at least, 
NJL-type models. Recent developments on this class of models \cite{pp} 
show them to be able to describe, in a physically transparent way, low energy 
$\pi -\pi$ scattering. This arises because Ward identities force the cancellation 
(for $m_\pi =0$) of all $\pi -\pi$ T-matrix contributions which are independent 
of the pion momenta. This observation (the Adler zero) 
implies a two-pion overlap kernel proportional to $p^2$.
This general observation, when taken together with the fact that pions have a size 
(a cutoff in momentum space), must then lead to overlap kernels numerically behaving as 
$p^2e^{-\beta p^2}$ with $\beta^{-1}$ playing the role of this cut-off. 
Other overlap kernels, with higher rank tensors in the relative $\pi -\pi$ momenta, may also exist. 
This requirement is model 
independent. It then happens that this type of kernels
is known, starting with the work of Van Beveren {\it et al.} \cite{doubling}, 
to hold, if strong enough, a light sigma. 
This is probably the common cause underneath the various independent calculations where a light sigma was found.

Similarly, we would also like to argue vacuum replication to be a general feature 
of low energy effective theories for hadron physics. This hope is reinforced by the map
which is 
known to exist between extended NJL models and $ChPT$ Lagrangians \cite{cahill}. However 
it remains to be seen whether this map is also able to inject the existing vacuum replication of NJL 
into $ChPT$. In any case it is clear that the Weinberg limit for the
pion-pion scattering \cite{We} is left untouched by the existence of such a vacuum replica. 
Indeed, the difference $\Delta\vp=\vp_0-\vp_1$ playing a crucial role for the theory of the
vacuum replica vanishes as $|\vec{p}|$ at small momenta $\vec{p}$. The amplitudes of physical
processes with the excitation of the replica are, roughly speaking, proportional to
$(\Delta\vp)^2$ or higher, so that they vanish at least as $\vec{p}^2$ when
$\vec{p}\to 0$, as compared to the amplitudes of the same processes without 
excitation of the replica. On the other hand, for finite momenta, the excitation of the vacuum 
replica does provide corrections to the standard formulae, which are interesting and 
instructive to extract and to study both theoretically and experimentally.

The Feynman rule of Eq.~(\ref{Gn}) together with Eq.~(\ref{DSeq}) constitute the central 
results of this work. They show that the existence of a replica is tantamount to 
the existence of a {\em new quark-quark force} going like (at BCS level)
$G(p)=nf(p_0)f(-p_0)(2\pi)^3\delta^{(3)}(\vec{p})$ and a quark energy renormalization,
$E_p\to E_p+\frac{n}{8}(\Delta\vp)^2$. This new interaction will add to sigma-exchange quark-quark 
forces in the intermediate energy range. Thus, we predict a new scalar
particle-like object which exists in the theory {\it together} with the 
\lq\lq standard" \cite{doubling} $\sigma$-meson. In contrast to the latter, the vacuum-induced
$\sigma$ has not a fixed mass, but rather a distribution given by the function
$f(\omega)$. The structure of this new particle in terms of the fermionic fields
is rather peculiar and has nothing to do with the naive scalar structure
$\bar\psi\psi$. Finally this new quark-quark force can be used to evaluate the decay of the replica 
into pairs of pions (and/or photons), as the lightest particles. This will be the subject 
of a future work.

\begin{acknowledgments} The authors are grateful to P. Bicudo and Yu. S. Kalashnikova for many fruitful discussions and for reading the manuscript and critical comments.
A. Nefediev would like to thank the
staff of the Centro de F\'\i sica das Interac\c c\~oes Fundamentais (CFIF-IST) for 
cordial
hospitality during his stay in Lisbon and to acknowledge the financial support of 
RFFI grants 00-02-17836, 00-15-96786, 01-02-06273 and INTAS grants OPEN 2000-110 and 
YSF 2002-49.
\end{acknowledgments}

\end{document}